# Transport Properties and Exponential $n$-values of Fe/MgB$_2$ Tapes With Various MgB$_2$ Particle Sizes

*P. Lezza, V. Abächerli, N. Clayton, C. Senatore, D. Uglietti, H.L. Suo, R. Flükiger*

Département de Physique de la Matière Condensée, Université de Genève, 24 Quai Ernest Ansermet, CH-1211 Genève 4 Switzerland

PAOLA LEZZA
Université de Genève
Département de Physique de la Matière Condensée
24, Quai Ernest Ansermet
CH-1211 Genève 4
Switzerland

E-mail paola.lezza@physics.unige.ch
Tel. 0041-22-379 66 69
Fax. 0041-22-781 09 80

**Abstract**

Fe/MgB$_2$ tapes have been prepared starting with pre-reacted binary MgB$_2$ powders. As shown by resistive and inductive measurements, the reduction of particle size to a few microns by ball milling has little influence on B$_{c2}$, while the superconducting properties of the individual MgB$_2$ grains are essentially unchanged. Reducing the particle size causes an enhancement of B$_{irr}$ from 14 to 16 T, while J$_c$ has considerably increased at high fields, its slope J$_c$(B) being reduced. At 4.2K, values of $5.3 \cdot 10^4$ and $1.2 \cdot 10^3$ A/cm$^2$ were measured at 3.5 and 10 T, respectively, suggesting a dominant role of the conditions at the grain interfaces. A systematic variation of these conditions at the interfaces is undertaken in order to determine the limit of transport properties for Fe/MgB$_2$ tapes. The addition of 5% Mg to MgB$_2$ powder was found to affect neither J$_c$ nor B$_{c2}$. For the tapes with the highest J$_c$ values, very high exponential *n* factors were measured: *n* = 148, 89 and 17 at 3.5, 5 and 10T, respectively and measurements of critical current versus applied strain have been performed. The mechanism leading to high transport critical current densities of filamentary Fe/MgB$_2$ tapes based on MgB$_2$ particles is discussed.

**Introduction**

$MgB_2$ is a promising superconductor [1] for cables, MRI and transformers. Moreover, conductor fabrication costs are lower compared to those of high temperature superconductors. Many efforts have already been made concerning the fabrication of superconducting $MgB_2$ wires and tapes by both in-situ and ex-situ methods [2]. In particular it has already been shown that Fe [3,4] and steel are the most suitable sheath materials to produce high $J_c$, powder-in-tube $MgB_2$ wires and tapes. It has been found that reducing the starting powder grain size leads to an increase of the critical current density $J_c$ and of the irreversibility field. The deformation and the conditions of the heat treatment have an influence on the texturing of the powder and on the transport properties [5], respectively.

In this paper further systematic developments and optimisation of the grinding process are presented. The enhancement of the critical current for finer powders has been confirmed. The production of sufficiently long tapes allowed transport critical current measurements under tensile strain.

**Experimental**

The powder-in-tube (P.I.T.) fabrication technique has been described elsewhere [2,6]. In this work we have analysed the effect of the initial $MgB_2$ powder size by ball milling it for different time intervals under a protected atmosphere to avoid oxygen inclusions enhancing their reproducibility. An iron tube is used in the P.I.T. process, which is filled with $MgB_2$ powder, swaged to 3.85 mm and drawn to 2 mm. As a last step the wire is deformed by flat rolling to obtain a 3.9 x 0.38 $mm^2$ tape. Currently a 920°C final heat treatment is performed for 0.5 hours (h).

**Results**

*1. Transport critical current densities $J_c$ versus applied field*

Two Fe/MgB$_2$ tapes have been prepared by ball milling the starting powder for 3 h (tape (a)) and 100 h (tape (b)) under Ar atmosphere. The powder distributions are shown in Fig. 1. The chemical analysis shows that after ball milling the oxygen content did not change in comparison to the initial content, 2 %. The process of ball milling the powder for 100 h instead of 3 h leads to an increase of the percentage of sub-micron grains, leaving the maximum diameter around 20 µm. In Fig. 2 the transport critical current $J_c(B)$ measurements at 4.2 K for the tapes (a) and (b) are shown. The results show that the highest $J_c(B)$ is obtained by grinding the powder for longer times i.e. with the smallest grain size. At 10 T the tape (a) reaches $J_c$ values of $= 3 \cdot 10^2$ A/cm$^2$ while the corresponding value for the tape (b) is higher than $10^3$ A/cm$^2$. This suggests that the connection between the grains and the surface available for the percolation of the current without MgO around the grains is enhanced. It is important to note that the tapes are very homogeneous, thus providing good reproducibility and the possibility to fabricate and measure long length of tape.

*2. Effect of Mg addition*

The powder prepared for the tape (b) has been doped with 5 % Mg after ball milling and a third tape has been fabricated (tape (c)) to compensate for the possible local Mg losses and possibly to bind the oxygen impurities at the surface of the original particles. The critical current densities $J_c(B)$ are also shown in Fig. 2. No significant variations were observed neither for $J_c$ nor for $B_{c2}$ (see section 5).

*3. Transport critical current densities at different temperatures*

Fig. 3 shows the critical current density versus applied field at different temperatures from 30 to 4.2 K for tape (b). At 4.2 K $J_c$ values above $10^3$ A/cm$^2$ and $4 \cdot 10^4$ A/cm$^2$ at 10 and 4.5 T, respectively, have been reached, while at 20 K the magnetic field has been ramped from 5.5 to 4 T with values of $4 \cdot 10^2$ A/cm$^2$ and $5 \cdot 10^3$ A/cm$^2$, respectively. It is interesting to compare the data in Fig. 3 with earlier values from bulk samples by Dhallé et al. [7]. For the high pressure sintered sample at 35 kbar at 950°C for 0.5 hours, the inductive critical current value was ≈ $10^4$ A/cm$^2$ at 3 T, 20 K. Extrapolation of our measurements at 20 K on tape (b) yields similar values. Moreover, it is also possible to make a comparison with the recent data from Serquis et al. [8] who prepared wires by the P.I.T. technique and treated them under hot hisostatic pressure (HIP). The $J_c$ of their wires are equivalent to that of tape (b). The main difference between the round wires submitted to the HIP process and the tapes is the degree of texturing which is inherently lower for round wires. It follows that the HIP process increases the densification of the powder, thus leading to an increased critical current density.

*4. n-values of 3 h and 100 h ball milled tapes*

For both tapes (a) and (b), transport critical current measurements allowed us to fit the *n*-value between the two criteria 0.1 µV/cm and 1 µV/cm in a log-log plot. In Fig. 4 the calculated *n*-values for the two tapes are shown; the circles and the triangles being tape (a) and (b), respectively. Tape (b) has very high *n*-values, being higher than 30 at magnetic fields lower than 9.5 T and reaching a maximum at 4.5 T of $n$ = 148 enhancing the possibility to use the conductors in persistent mode operation'. The spread of the data is probably due to the poor thermal stabilisation of the tapes which leads to increased noise in the superconducting-to-normal state transitions.

*5. Third harmonic measurements and irreversibility field $B_{irr}$*

For tape (b) and (c) the upper critical field $B_{c2}$ has been measured and the results are shown in Fig. 5; the values are similar to those previously reported [9].

The irreversibility field $B_{irr}$ can be obtained from transport or inductive measurements [2]. A third method takes into account the onset of the third harmonic susceptibility in the presence of an applied magnetic field $H_{dc} \gg H_{ac}$ [10].

The onset of the third harmonic response results from the crossover from a flux flow dominated regime, characterized by the absence of harmonic signal to a pinning dominated one, corresponding to a non zero critical current. For both the (a) and (b) tapes, measurements of third harmonic signal have been performed in a 9 T magnet, and the data are shown in Fig. 5. The extrapolation to 0 K leads to high $B_{irr}$ values, above 15 T for tape (a) and above 16 T for tape (b). Moreover, in Fig. 3 it is possible to take into account a criterion of 10 A/cm$^2$ and to extrapolate the irreversibility field $B_{irr}$ thus confirming the measurements performed with the third harmonic signal. The enhancement of $B_{irr}$ leads to values that are not far from $B_{c2}$. This suggests that the enhancement of $B_{c2}$ must be achieved by other methods such as doping or irradiation [11].

*6. Effect of tensile strain on critical current density $J_c$*

Measurements of critical current $I_c$ versus tensile strain have been performed using a modified Walters Spring whose sample holder allows us to measure the critical current on an 80 cm length of tape. Strain gauges are applied to the sample to measure the critical current under compressive or tensile strain. In Fig. 6 data for tape (b) are shown at 7 and 9 T under tensile strain. At 9 T the critical current goes up by 10 % until 0.26 % strain. Increasing the strain to 0.31 % leads to a reduction of $I_c$ below the zero strain value. If the strain is released at this point, $I_c$ is no longer recovered because the filament is broken. At 0.36 % strain the critical current is zero. The behaviour is clearer at 7 T; the tape is extended until 0.27 % strain which enhances the critical current by 6% in comparison to the critical current measured at zero strain. At 0.32 % strain the critical current is 20 % lower and the tape is broken.

**Conclusions**

$MgB_2$ tapes have been prepared with pre-reacted powder to study the behaviour of the superconducting properties. It has been found that the enhancement of the transport properties is due to the reduction of the powder grain size avoiding the oxygen content to raise. The Mg addition did not lead to significant variations in the superconducting and transport properties of the conductors. The critical current density $J_c$ and the irreversibility field $B_{irr}$ have been raised from $3 \cdot 10^2$ to $10^3$ A/cm$^2$ and from 14 to 16 T, respectively thanks to the grinding of the starting powders. Measurements of the critical current density versus tensile strain have been performed and an enhancement of the critical current density by 5 % and by 10 % at 7 and 9 T, respectively for 0.25 % of applied strain was observed. At higher strain the filament was broken. In this case, the degradation of the critical current and *n*-values confirms the pre-stress model; the behaviour is similar to that of $Nb_3Sn$ conductors. The tapes were very homogeneous indicating that long lengths can be produced. Further improvement of pinning properties is in progress to raise the upper critical field $B_{c2}$ and to reach the thermal stability of the conductors as well as the deep understanding of the Mg additions.

**Figure Captions**

Fig. 1: Grain size distributions for the two ball milled powders 3 hours (a) and 100 hours (b).

Fig. 2: Critical current densities $J_c(B)$ for the tapes fabricated using the 3 hours (b), 100 hours (b) ball milled powders and 100 hours ball milled powders with 5 % Mg addition (c).

Fig. 3: Critical current densities $J_c(B)$ at different temperatures for the tape (b) compared with a bulk sample by Dhallé et al. [7] and a HIPed wire by A. Serquis et al. [8].

Fig. 4: *n*-values for the tape (a) (circles) and (b) (triangles) calculated as the slope in the log-log plot between the two criteria 0.1 µV/cm and 1 µV/cm.

Fig. 5: Irreversibility field $B_{irr}$ for the two tapes (a) and (b) measured using the onset of the third harmonic (an example of the measurement in the inset) and upper critical field $B_{c2}$ for the tape (b) and (c) measured by the resistive transition criterion.

Fig. 6: Critical current densities $J_c(B)$ at 7 and 9 T for the tape (b) as a function of the applied tensile strain.

Figure 1

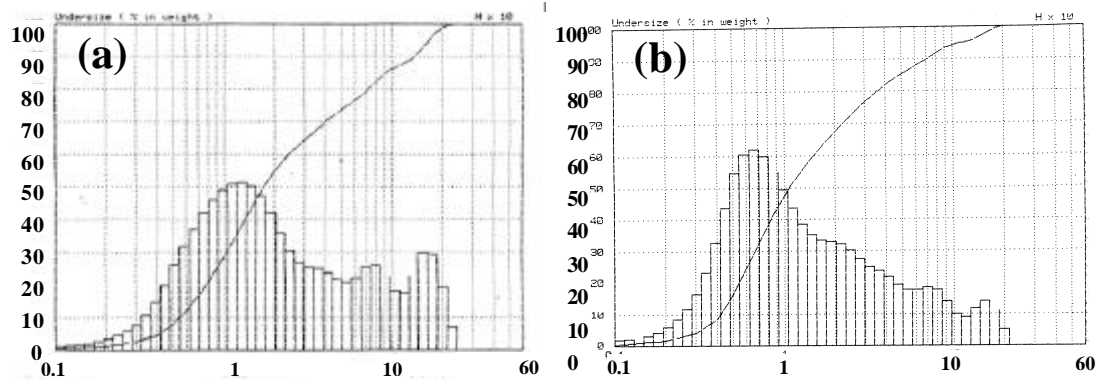

Figure 2

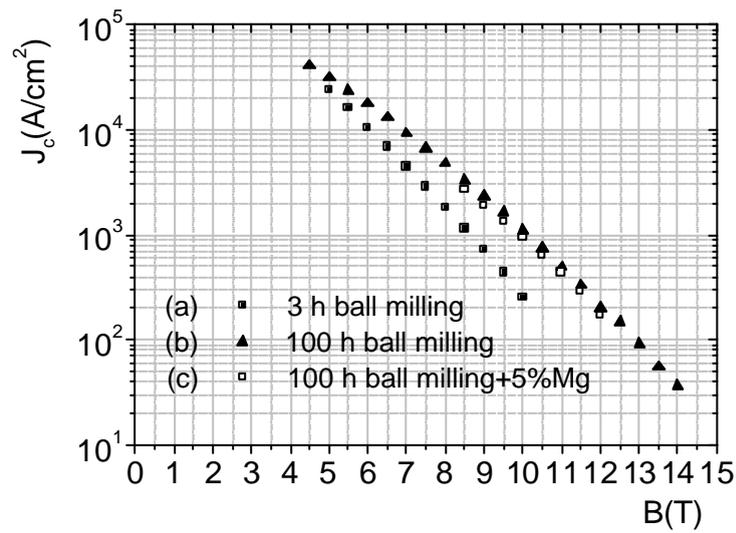

Figure 3

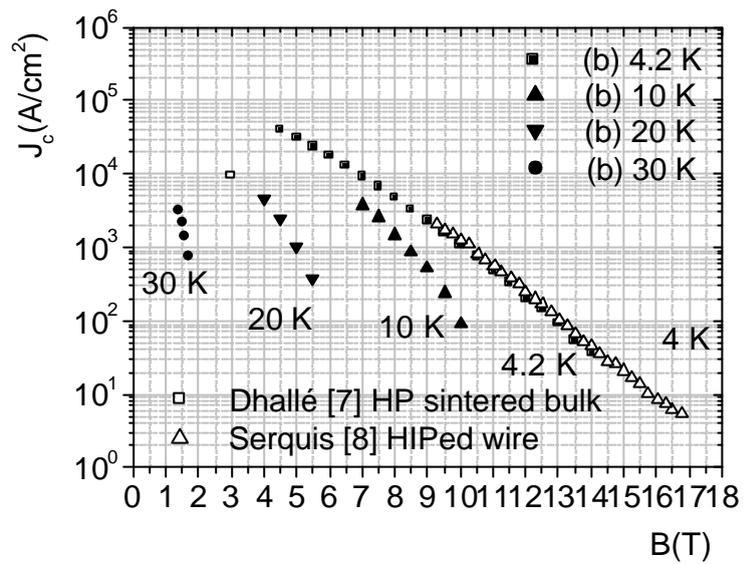

Figure 4

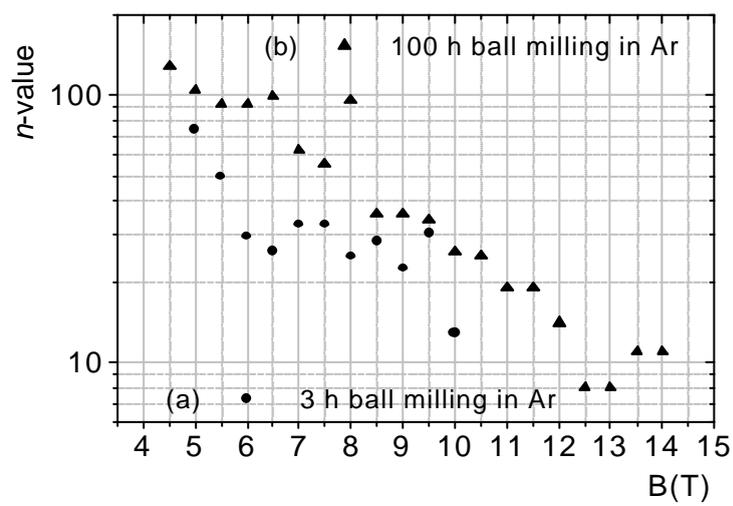

Figure 5

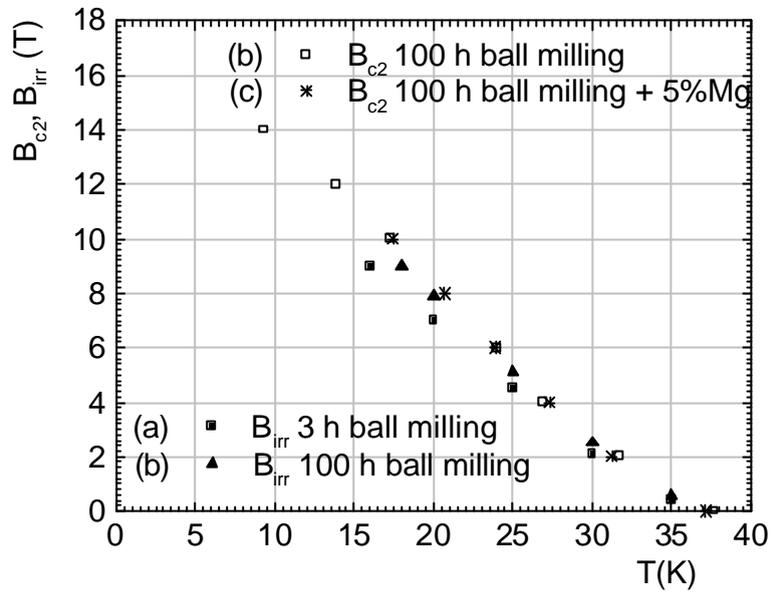

Figure 6

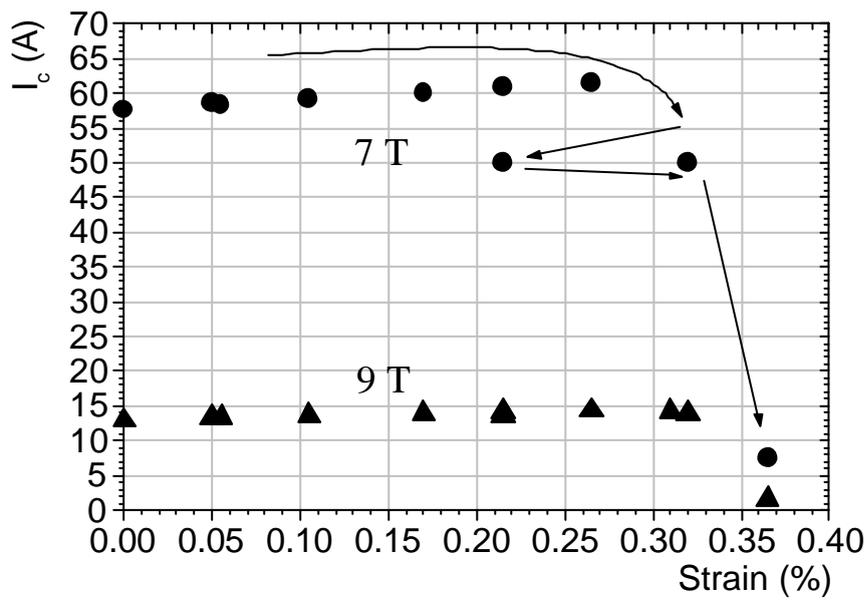